# High refractive index composite materials for THz waveguides: trade-off between index contrast and absorption loss

Bora Ung, Alexandre Dupuis, Karen Stoeffler, Charles Dubois and Maksim Skorobogatiy[a]

*Engineering Physics Department, Ecole Polytechnique de Montréal, C.P. 6079, succursale Centre-Ville Montreal, Québec H3C3A7, Canada*

Polymer compounds from titania-doped polyethylene are fabricated and their linear optical properties characterized by THz-TDS. We show that high concentration of dopants not only enhances the refractive index of the composite material, but also can dramatically raise its absorption coefficient. We demonstrate that the design of Bragg reflectors based on lossy composite polymers depends on finding a compromise between index contrast and corresponding losses. A small absorption value is also shown to be favorable, compared to an ideal lossless reflector, as it enables to smooth the transmission passbands. Transmission measurements of a fabricated hollow-core Bragg fiber confirm simulation results.

The development of THz waveguides is motivated by the need for efficient delivery of terahertz radiation to enable remote terahertz sensing, imaging and spectroscopy applications. The chief obstacle for realizing low-loss THz waveguides is that most materials exhibit large absorption losses inside the terahertz spectrum. A solution found to partly overcome this problem is to design either hollow-core or highly porous waveguides, or subwavelength solid-core waveguides, in all of which a substantial portion of the optical power is guided in dry air. In line with this scheme, several types of microstructured fibers have been designed from low-loss polymers: photonic crystal fibers [1, 2], porous fibers [3-5], hollow Bragg fibers [6] and anti-resonant reflecting optical waveguides (ARROW) [7]. Our study here focuses on hollow-core Bragg fibers which enable to tune almost arbitrarily the width and amplitude of the band gaps by controlling the index contrast and thicknesses of the dielectric layers constituting the 1D periodic reflector. In this regard, there are strong incentives in developing of high-refractive index materials for enabling THz optical components and waveguides.

In this Letter, we demonstrate how increasing the concentration of subwavelength-size micro-particles inside a polymer elevates the refractive index of the dielectric layers; but also significantly raises the absorption losses of the composite material. We present experimental optical characterizations performed on doped-polymer films and on a hollow-core Bragg fiber fabricated by our group. Our analysis reveals that for the design of THz Bragg fibers or filters, a balance between the elevation of the refractive index and the losses must be found; or else overly large losses may completely suppress the band gaps of the fiber.

For our numerical calculations, we assumed a perfectly circularly symmetric Bragg fiber composed of 14 bi-layers [Fig. 1(a)] created by the periodic stacking of high-index doped-PE and low-index undoped-PE films of respective thicknesses $d_H$=135 μm and $d_L$=100 μm, and core diameter $d_{core}$ = 6.63 mm. Due to the high fragility of the doped layers, the fabrication process currently remains challenging, hence the produced Bragg fiber [Fig. 1(b)] has fewer bi-layers: 5.

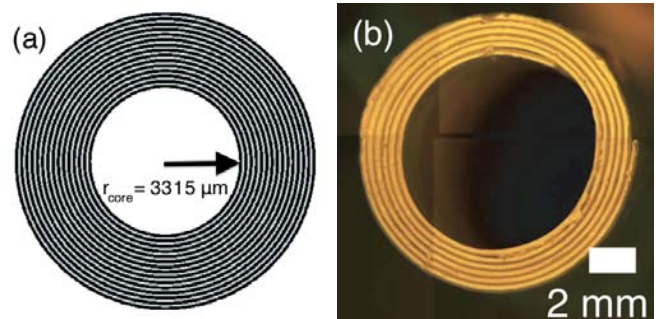

FIG. 1. (Color online) Hollow polymer Bragg fiber geometry: the reflector is made of alternating high-index $TiO_2$-doped PE layers and low-index pure-PE layers. (a) Structure with 14 bi-layers implemented in transfer-matrix calculations. (b) Bragg fiber with 5 bi-layers fabricated in-lab by film extrusion, hot pressing and subsequent coiling around a mandrel.

A large increase in the terahertz refractive index has recently been demonstrated by incorporating high-index subwavelength-size particles within a host polymer [8-10]. Here we similarly doped linear low-density polyethylene (PE) with high-index rutile titanium dioxide ($TiO_2$) micro-particles at varying concentrations so as to control the increase in the values of the refractive index and absorption coefficient of the polymeric compound. THz-TDS transmission measurements performed on the fabricated $TiO_2$-doped and undoped PE films showed that the refractive index of the polymer compounds remained practically constant throughout the 0.1-2.0 THz range [inset of Fig. 2]. The refractive index of the PE-$TiO_2$ polymeric compound was modeled with Bruggeman's effective medium approximation [10]:

[a] URL: http://www.photonics.phys.polymtl.ca  Electronic mail: maksim.skorobogatiy@polymtl.ca

$$1 - f_v = \frac{\varepsilon_p - \varepsilon_m}{\varepsilon_p - \varepsilon_h} \sqrt[3]{\frac{\varepsilon_h}{\varepsilon_m}} \quad (1)$$

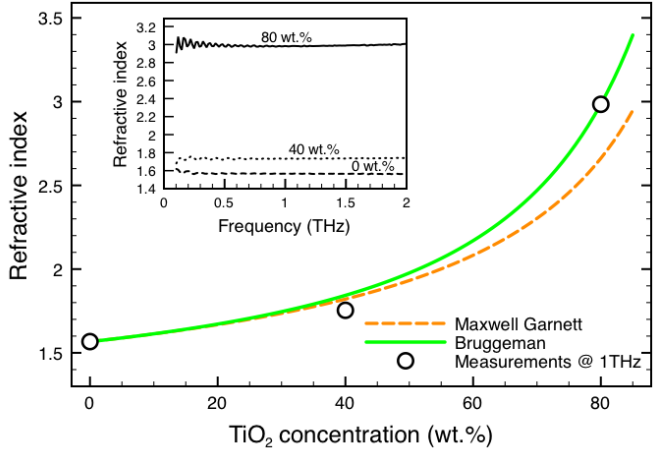

FIG. 2. (Color online) Refractive index of PE-based polymer compound as a function of weight concentration of $TiO_2$ doping: Bruggeman model fit (solid line), and measurements at 1 THz (circles). Inset: THz-TDS measurements of the refractive index for pure PE (dashed line), 40 wt.% (dotted line) and 80 wt.% (solid line) $TiO_2$ doping.

where $f_v$ is the volumetric particle fraction ($0 < f_v < 1$) of doping particles, and $\varepsilon_p$, $\varepsilon_h$ and $\varepsilon_m$ denote the respective permittivities of the dielectric particles, host and mixture. While the Maxwell-Garnett (MG) effective medium model is only valid for dilute volume concentrations ($f_v < 15$ vol.%) of spherical particles within a host material [10]; the Bruggeman mixing model extends the validity domain of the MG approach for high concentration of dopants by taking into account the distinct interactions of the two-phase composite medium. We measured for the host pure PE: $\varepsilon_h = 2.455$ (or $n_h = 1.567$). Since no terahertz permittivity data for our rutile $TiO_2$ grade was available, a best-fit value of $\varepsilon_p = 39.5$ for the $TiO_2$ particles was extracted by fitting Eq. (1) with the measurements of the mixture's refractive index (Fig. 2). The relation $f_v = f_w / (f_w + (1 - f_w)\rho_p/\rho_h)$, where $\rho_p$ and $\rho_h$ are the densities of each compound, is useful for converting a weight fraction $f_w$ of particles into a volume fraction $f_v$. The density of PE polymer provided by the manufacturer (Nova Chemicals: SCLAIR polyethylene FP120-A grade) is $\rho_h = 0.920$ g/cm$^3$ and that of rutile $TiO_2$ (DuPont: Ti-Pure R-104) is $\rho_p = 4.2743$ g/cm$^3$. Due to the large difference in densities, it follows that a high weight fraction $f_w = 80$ wt.% of $TiO_2$ relates to a much lower volume fraction: $f_v = 46.3$ vol.%. Using the latter conversion relation and Eq. (1), the Bruggeman model gives refractive indices in good quantitative agreement with experimentally measured values [see Fig. 2].

As shown on Fig. 3, the material absorption loss exhibited a quadratic increase with frequency in the case of doped polymers. The $TiO_2$-doped PE power loss coefficient ($\alpha_m$) in cm$^{-1}$ was empirically modeled as a function of the volumetric doping fraction ($f_v$) and input frequency ($\nu$) via the following second-order polynomial fit:

$$\alpha_m = f_v (a_1 \nu^2 + a_2 \nu + a_3) + a_4 \quad (cm^{-1}), \quad \nu = (THz) \quad (2)$$

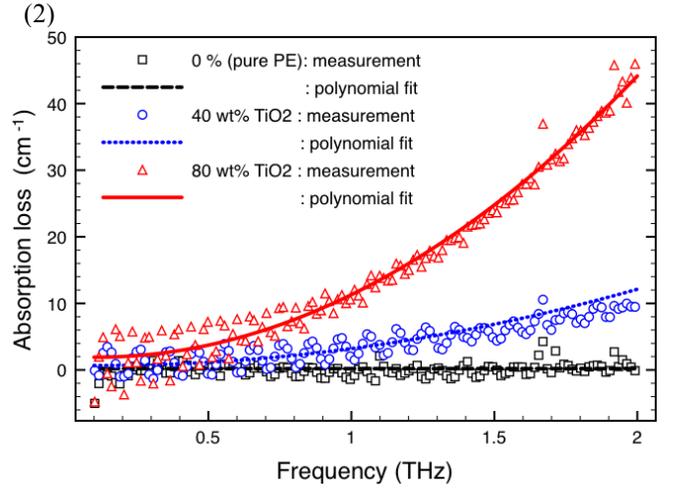

FIG. 3. (Color online) Material power absorption loss of PE-based polymer compound as a function of frequency (THz) for various levels of $TiO_2$ doping concentrations: 0 wt.% (pure PE), 40 wt.% and 80 wt.%. Squares, circles and triangles denote THz-TDS measurements while the dashed, dotted and solid curves represent the corresponding polynomial fits [see Eq. (2)].

For our $TiO_2$-PE compound, the coefficients are: $a_1 = 25.5$, $a_2 = -5.5$, $a_3 = 4.0$ and $a_4 = 0.2$. Coefficient $a_4$ actually defines the base absorption of undoped PE ($f_v = f_w = 0$) which oscillates (due to Fabry-Pérot effects in planar samples [9]) about an average approximate value of 0.2 cm$^{-1}$ in the range 1.0–2.0 THz.

Since a linearly polarized source is used in our setup, only non-azimuthally polarized modes may be excited in the fiber. The attenuation coefficient of the lowest-order fundamental $HE_{11}$-like mode was computed with a semi-analytic transfer-matrix mode solver using a high refractive index $n_m = 2.985$ corresponding to $f_w = 80$ wt.%. To investigate the effect of material losses on the band gaps, we gradually increased the losses of the reflector ($\alpha_{layers}$) through $\alpha_{layers} = f_{layers}\alpha_m$ where $f_{layers}$ specifies the fraction (0%, 10%, 30% up to 100%) of the layers' nominal loss $\alpha_m$ defined by Eq. (2).

The cyan dotted curve in Fig. 4 corresponds to a doped Bragg fiber with *no* absorption losses ($f_{layers} = 0\%$) in the bi-layers. This idealized configuration reveals 10 distinct band gaps inside the THz range, with the widest being the fundamental band gap located at 0.3 THz. To shift the fundamental band gap to the 1 THz frequency, one would need to scale down the size of dielectric layers by a factor of 3. For frequencies inside a band gap, the light is tightly confined within the central lossless core and thus guided with very low attenuation (as seen on Fig. 4). Outside of the band gaps however, a substantial fraction of guided power leaks out into the stack of dielectric layers. The appearance of oscillating fringes for these out-of-bandgap regions can be simply explained by viewing the stack as a single dielectric capillary tube (i.e. a thick ARROW fiber), of average refractive index, that acts as a Fabry-Pérot resonator inside of which light rays undergo multiple reflections.

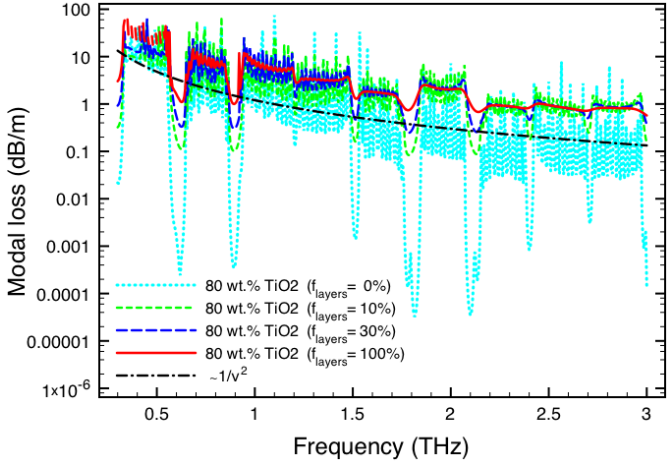

FIG. 4. (Color online) Fundamental mode loss in doped-PE/PE hollow-core polymer Bragg fibers (14 bi-layers) as a function of input frequency and for various levels of fractional nominal loss ($f_{layers}$) inside the dielectric layers.

Figure 4 indicates that the elevation of losses in the reflector via doping essentially flattens the features of the transmission spectrum: first it smoothes out the oscillating fringes due to "ARROW-type" effects in out-of-bandgap regions, and it elevates the minimum attenuation loss inside the band gap regions. Another (beneficial) effect is the removal of attenuation spikes in the middle of some band gaps. It is important to note that the flattening rate of oscillating fringes is greater than that of the band gaps. In the case of $f_{layers}$ = 30%, the dynamic range of fringes is suppressed (compared to the lossless case) by a factor of more than 6; while that of band gaps is reduced by approximately 4.5.

The solid red curve in Fig. 4 illustrates the extreme case of very high losses ($f_{layers} = 100\%$) for which both band gaps and oscillating fringes are strongly suppressed; here instead, light guidance mainly occurs via pure Fresnel reflection from the first dielectric layer. This last result was confirmed by our experimental setup [see Fig. 5]. Especially at frequencies above 1.5 THz, the very high losses eliminate any resonant behavior; and what mostly remains is the relation between radiation losses and frequency [11]: $\alpha_{rad} : 1/(r_{core}^3 v^2)$ depicted by the dashed-dotted curve. In more detail, the total guided mode attenuation is determined by the contributions of the *material absorption* [Eq. (2)] and the *radiation losses* ($\alpha_{rad} = b_4/v^2$) as follow: $\alpha_{mode} = \alpha_m \cdot \alpha_{rad} = b_1 + b_2 v^{-1} + b_3 v^{-2}$ where $b_1$, $b_2$, $b_3$ and $b_4$ denote scalar coefficients.

The precise spiral asymmetric shape of the fiber (including the inner/outer step defect shown in inset of Fig. 5) was implemented for the finite-element (FEM) simulations. The FEM calculations [dotted curve in Fig. 5] of the fundamental mode propagation losses indicates that the hollow-core's broken symmetry created by the small inner step results in a slight ( 0.05 THz) blue-shifting of some peaks of radiation loss located near the blue end of the THz range (where the dimension of light wavelength becomes closer to that of the step defect) when compared to a perfectly circular symmetric Bragg fiber [dashed curve in Fig. 5].

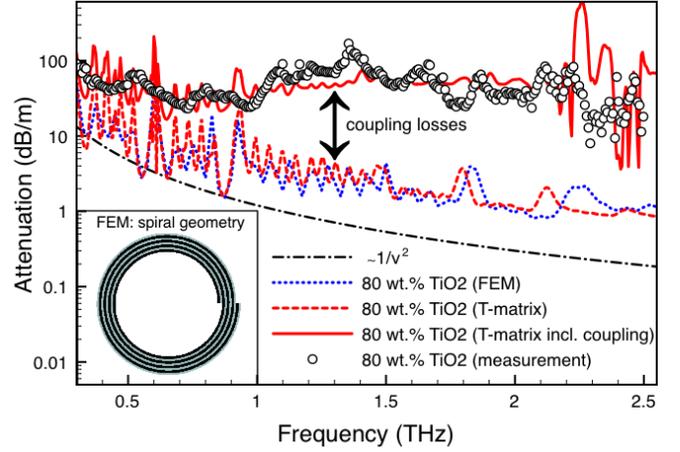

FIG. 5. (Color online) Fundamental mode loss (dB/m) in 80 wt.% doped Bragg fiber computed by transfer matrix method (blue curve) with perfect circular symmetry, and FEM (red curve) with the asymmetric spiral shape (shown inset). THz-TDS measurements of total attenuation (open circles), and theoretical calculations including coupling losses (solid curve).

The THz-TDS transmission measurements of the total signal attenuation – including (in-and-out) coupling losses via the parabolic mirrors of the setup – are denoted by the open circles in Fig 5. An estimation of the coupling losses was obtained by the overlap integral of the input Gaussian field over that of the calculated propagating $HE_{11}$ mode field, at each frequency. The results of simulations taking into account coupling losses (solid line in Fig. 5) and the measurements agree on the scale and overall behavior of the attenuation spectrum. Because of its large diameter, the Bragg fiber is highly multimode and it has been shown to affect significantly the transmission spectrum due to coupling in higher-order modes and inter-modal interferences [12]. Also, the cross-section may not fully retain its input facet shape, and some small defects might appear along the fiber length. The latter remarks generally explain the visible discrepancies between measurements and simulations.

In summary, we experimentally and theoretically investigated high refractive index THz materials based on polymeric compounds. Our modal analysis of hollow-core Bragg fibers reveals that large absorption losses induced by the dopants can effectively destroy the passbands (i.e. band gaps). We argue that a proper balance between the index contrast (controlled via doping concentration), and the related material losses, enables to create wide and smooth band gaps in the THz transmission spectrum. These latest findings are relevant for the design of THz waveguides and filters based on composite materials.